\documentclass[conference]{IEEEtran}
\ifCLASSINFOpdf
 \usepackage[pdftex]{graphicx}
\DeclareGraphicsExtensions{.pdf,.jpeg,.png}
\else
\fi

\usepackage{caption}

\begin{document}
%
\title{Privacy and Anonymity}

\author{\IEEEauthorblockN{Adrian Yanes}
\IEEEauthorblockA{Aalto University - School of Electrical Engineering\\
Department of Communications and Networking (ComNet)\\
\textit{Challenged Networks S-38.3455 P} \\
adrian.yanes@aalto.fi}}


%


\maketitle
\thispagestyle{plain}
\pagestyle{plain}
\begin{abstract}

Since the beginning of the digital area, privacy and anonymity have been
impacted drastically (both, positively and negatively), by the different
technologies developed for communications purposes. The broad
possibilities that the Internet offers since its conception, makes it a
mandatory target for those entities that are aiming to know and control
the different channels of communication and the information that flows
through. \\
In this paper, we address the current threats against privacy and
anonymity on the Internet, together with the methods applied against them. 
In addition, we enumerate the publicly known entities behind those threats and their
motivations. Finally, we analyze the state of the art concerning the
protection of the privacy and anonymity on the Internet; introducing future lines of research.
\end{abstract}


%
\IEEEpeerreviewmaketitle

\section{Introduction}

Despite the familiarity of the concepts, privacy and anonymity are
commonly misunderstood within the digital context. In this paper, we
refer to these concepts as foundations of the Common Criteria for
Information Technology Security Evaluation (CC)
standard~\cite{iso15408}. Therefore, we strictly refer to the
implications of these concepts within the technological jargon, with a
deep emphasis on the Internet. We make the following contributions:

\begin{enumerate}

\item {We provide a precise definition of the research terms,
  privacy and anonymity, and their implications.}

\item {We analyze the main threats against privacy and anonymity and
  their originators.}

\item {We share a resume of the state of art in terms of the
  protection of the privacy and anonymity on the Internet.}

\end{enumerate}

The remainder of the paper is organized as follows. In Section 2 we analyse
the main threats against privacy and anonymity and their originators.
In Section 3, we show the current methodologies applied to gain
control over the privacy and anonymity of the users. We enumerate the
current technologies available to protect privacy and anonymity on the
Internet in Section 4. In Section 5 we introduce the proposed improvements over the state of the art.
We present our conclusions in Section 6.


\subsection{Privacy}
In the past few decades there have been several debates about the precise
definition of privacy. The Universal Declaration of Human Rights~\cite{community1948universal} in both article 12 and 19, references the concept as a human right:

\textit {\textbf{Article 12}: No one shall be subjected to
  arbitrary interference with his privacy, family, home or
  correspondence, nor to attacks upon his honour and reputation [...]}

\textit {\textbf{Article 19}: Everyone has the right to freedom of
  opinion and expression; this right includes freedom to hold opinions
  without interference and to seek, receive and impart information and
  ideas through any media and regardless of frontiers.}

Other authors refer to privacy as \textit{``the
  claim of individuals, groups, or institutions to determine for
  themselves when, how, and to what extent information about them is
  communicated to others.''}~\cite{westin1970privacy}

According with the CC standard~\cite{iso15408}, privacy involves
\textit{``user protection against discovery and misuse of identity by
other users''}. In addition, the CC standard~\cite{iso15408} defines the
following requirements in order to guarantee privacy:

\begin{itemize}
  \item Anonymity
  \item Pseudonymity
  \item Unlinkability
  \item Unobservability
\end{itemize}

Thus, we consider the definition of privacy as a framework of
requirements that prevents the discovery and identity of the user.

\subsection{Anonymity}

As stated before, anonymity is intrinsically present in the concept of
privacy. Nevertheless, anonymity refers exclusively to the matters
related to the identity. The CC standard~\cite{iso15408} defines
\textit{``[anonymity] ensures that a user may use a resource or service
  without disclosing the user’s identity. The requirements for anonymity
  provide protection of the user identity. Anonymity is not intended to
  protect the subject identity. [...] Anonymity requires that other
  users or subjects are unable to determine the identity of a user bound
  to a subject or operation.''}~\cite{iso15408}~\cite{anon_terminology}

Accordingly, we consider the definition of anonymity as the property
that guarantees user's identity from being disclosed without consent.

\subsection{Other concepts involved in privacy \& anonymity}

\subsubsection{Pseudonymity}

Notwithstanding, the use of anonymity techniques can protect the user
from revealing their real identity. Most of the time there is a
technological requirement necessary to interact with an entity, thus,
such entity requires to have some kind of identity. The CC~\cite{iso15408}
claims \textit{[pseudonymity] ensures that a user may use a resource or service without disclosing its user identity, but can still
be accountable for that use}.

\subsubsection{Unlinkability}

In order to guarantee a protection of the user's identity, there is a
need for unlinkability of the user's activities within a particular
context. This involves the lack of information to distinguish if the
activities performed by the user are related or not.

\subsubsection{Unobservability}

The CC standard~\cite{iso15408} refers to this concept as
\textit{``[unobservability], requires that users and/or subjects cannot
    determine whether an operation is being performed.''}. Other
authors claim that unobservability should be differentiated from the
undetectability~\cite{anon_terminology}. The reasoning behind this, claims
that something can be unobservable, but can still be detected. In this paper
we refer to unobservability as the property that guarantees the
impossibility to distinguish if something exists or not.

\section{Background}

Historically, the control of the communications and the flow of
information, are mandatory for any entity that aims to gain
certain control over the society. There are multiple entities with such
interests: governments, companies, independent individuals, etc. Most
of the research available on the topic claims that the main originators of
the threats against privacy and anonymity are governmental institutions
and big corporations\cite{taxo}. The motivations behind these threats
are varied. Nevertheless, they can be classified under four
categories: social, political, technological and economical. Despite
the relation between them, the four categories have different backgrounds.

\subsection{Social \& political motivations}

The core of human interaction is communication in any
form. The Internet has deeply impacted in how social interaction
is conducted these days. The popularity and facilities that the Internet
offers, makes it a fundamental asset for the society. Currently, it is estimated that
there are more than \textbf{2.7 billion} individuals users of the Internet
in the world\cite{ICT}.

\begin{figure}[h]
  \centering
  \captionsetup{justification=centering}
  \includegraphics[width=8cm]{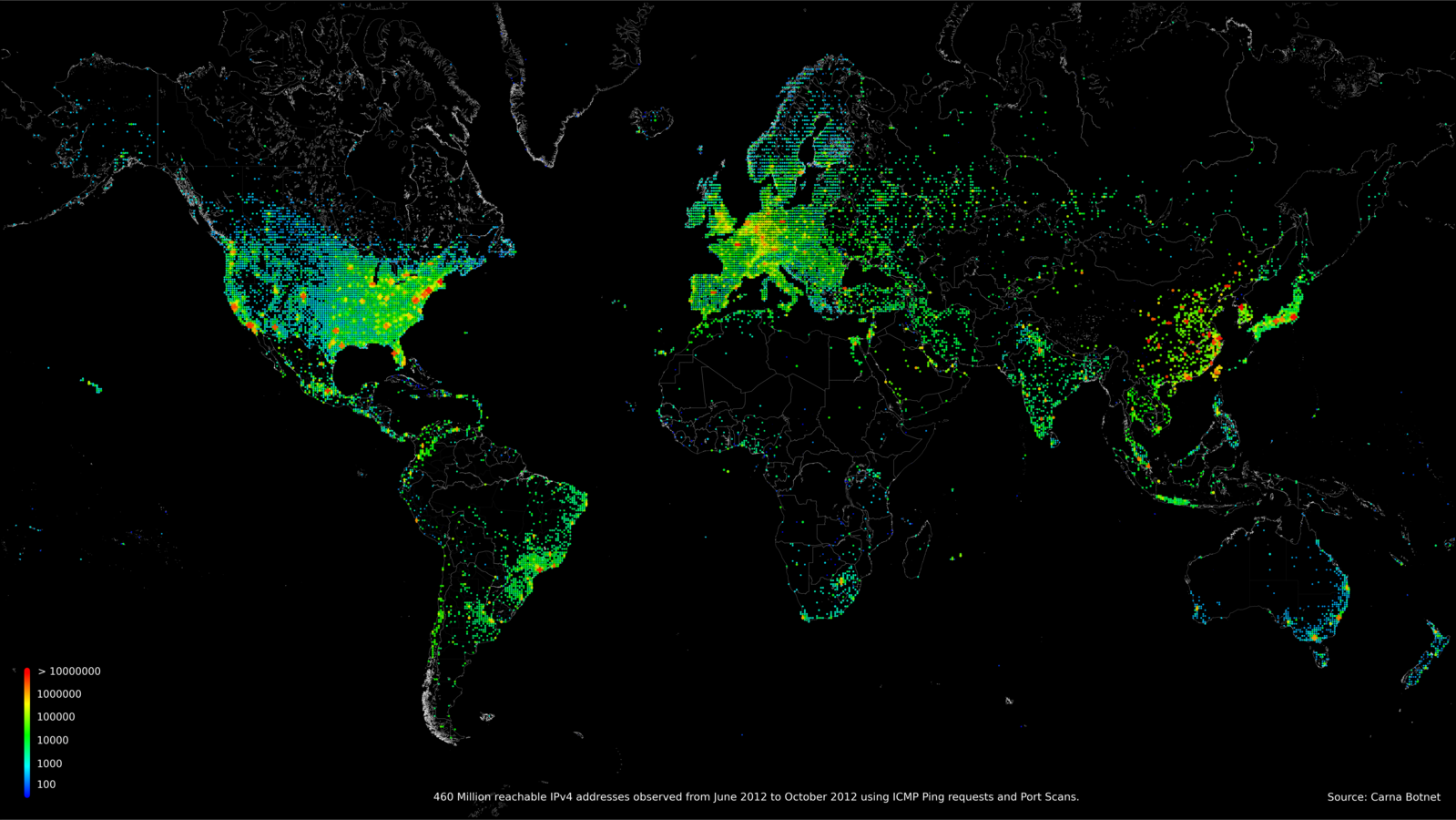}
  \caption{~460 Million IPv4 addresses on world map (2012)\cite{CENSUS}}
\end{figure}

Any entity that gains any level of control over this massive exchange
of information, implicitly obtains two main  advantages: the capability to
observe the social interaction without being noticed (hence, being
able to act with certain prediction), and the possibility
to influence it. Privacy and anonymity are the core values against
these actions. Nevertheless, several authoritarian regimes implemented
diverse mechanisms for the dismissal of both, privacy and anonymity.

Many of the authors
highlight that the foundations of these threats are most often
motivated for ideological reasons\cite{taxo}, thus, in countries in
which free of speech or political freedom are limited, privacy and
anonymity are considered an enemy of the state. In addition, national
defense and social ``morality'' are also some of the main arguments
utilized when justifying actions against privacy and anonymity\cite{ACCESS}.

According to the OpenNet Initiative\cite{OPENNET}, there are at
least \textbf{61 countries} that have implemented some kind of
mechanism that negatively affects privacy and anonymity. Examples of these countries are
well-known worldwide: China\cite{GFC}, Iran\cite{IRAN}, North Korea and Syria, among
others. In addition, recent media revelations\footnote{Global
  surveillance programs: PRISM(UK), Tempora(NSA), Muscular (NSA)} shown that several
mechanisms that are negatively affecting privacy and anonymity have been
implemented in regions such as the U.S. and Europe.

\subsection{Technological issues}

There are some cases in which the threats against privacy and
anonymity occur due to the lack of proper technology. Sometimes these
threats can occur unintentionally. An example of this are bugs in
software that are not discovered and somehow reveal information about
the identity or data of the users. Also, misconfigured Internet services that do
not use proper encryption and identity mechanisms when offering
interaction with their users. Certain techniques utilized by
the ISPs can lead to situations in which the user's data and identity gets
compromised even if the ISPs' intentions are focused on bandwidth
optimization. Finally, non-technological educated users can be a
threat to themselves by unaware leaking their identity and data
voluntarily but unaware of the repercussions (e.g. usage of social networks, forums, chats, etc).

\subsection{Economical motivations}

As stated previously, the impact and penetration of the Internet in 
modern society affects almost every aspect of it, with a primary use of it  
for commercial/industrial purposes. There are multiple economical
interests that are related directly to the privacy and anonymity of the
users. Several companies with Internet presence take advantage of
user's identity in order to build more successful products or to target a
more receptive audience. Lately, the commercialization of user's data has
proven to be a profitable business for those entities that have the
capability of collecting more information about user's behavior. In
addition, due to the popularity of Internet for banking purposes,
the privacy and anonymity of the users are a common target for malicious
attackers seeking to gain control over user's economical assets.

\section{Mechanisms used to interfere privacy and anonymity}

In the past few decades, a big technological and economical investment
has been made by some entities seeking to gain some control over the
privacy and anonymity on the Internet. Nevertheless, there are not too
many commercial technologies developed with this particular
purpose\cite{taxo}; forcing the development of custom solutions
adapted to the particular use case/target. Depending of the
originator, the technology utilized can have different types of
targets, the most common being nodes and individual users on the
Internet. The categorization of the technology falls in two areas:
hardware and software solutions. The devices utilized for these
targets are commonly denominated Deep Packet Inspection (DPI)
devices. Their main use is the classification of the network traffic,
together with the inspection of packet headers and payloads. Note that
most of these devices are not manufactured to directly target the privacy
and anonymity of the users. Instead, they have a generalist
purpose (commonly associated to routing and QoS
purposes). Nevertheless, these devices can be configured in ways that
the privacy and anonymity get affected.
 
We will analyze the different threats using the OSI model\cite{OSI} as
reference, highlighting the most common compromises on terms of privacy and
anonymity across the different layers.

\subsection{Physical Layer}

Any possible threat on the physical layer implies direct access to the
hardware involved in the network. Most common artifacts utilized are
known as network taps: devices that provide access to the data flow of
the network once they are attached to it. These devices can be used
either to monitor the network silently (\textit{sniffing}) or to
redirect the full traffic of the network to a different
node. Implicitly, any observation of the network traffic involves the
possibility of affecting the privacy and anonymity of the traffic.

\subsection{Data link layer}

Most of the hardware solutions targeting the data link layer focus on
the media access control (MAC) sublayer. It is in the MAC sublayer
that certain filters can be implemented. The addressing of the
destinations occurs in this layer, which are unique. This is the
first control point for those entities that are aiming to obtain some
information concerning user's identity, data flows and destinations. Note,
the property of uniqueness of the MAC address and the current
standard~\footnote{Referring here to the assignation of MAC address
  schemes by manufacturer}, allows the categorization of the devices
present in the network. This can lead to a premature identification of
the user (by manufacturer/model).

\subsection{Network layer}

Due to the nature of the Internet protocol suite, and the use of Internet Protocol (IP)
as its core; the network layer plays a crucial role while interfering
with the privacy and anonymity of the network traffic. Most of the DPI
hardware focused on the network layer targets the inspection of the
IP packets. The IP packets contain all of the relevant information
required for routing purposes. Therefore, it is a mandatory target
when aiming to reveal the identities of the users. The analysis is
performed mainly on the IP header; more specifically, on the source
address, destination address and protocol type. This data can be
utilized with different malicious purposes; the analysis and
processing of IP headers can lead to the direct identification of the
user involved in the communication. 

\begin{figure}[h]
  \centering
  \captionsetup{justification=centering}
  \includegraphics[width=8cm]{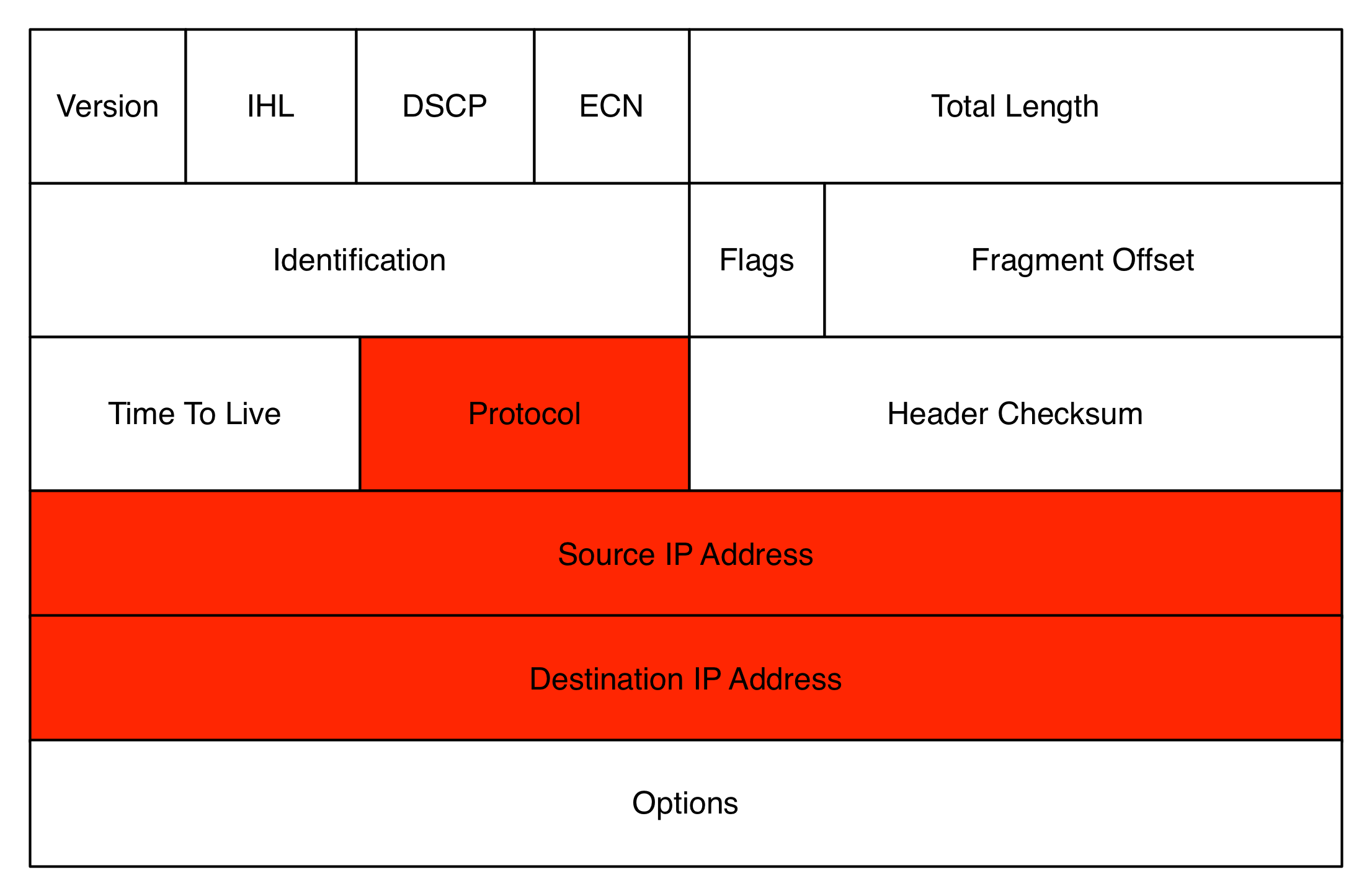}
  \caption{IPv4 header fields commonly targeted (red)}
\end{figure}

Furthermore, statistical analysis can
be performed to determine the population of users even if they are
using encrypted protocols over the IP\cite{SWARM}. Finally, the
identification of the protocol utilized, provides enough information
to perform further analysis of the data that the IP packet is
transporting. This makes the network layer a critical asset for
inspecting the network traffic. It is in this layer where
some of the entities previously mentioned implement heavy traffic
analysis and/or filtering.

\subsection{Transport layer}

The Transport Control Protocol (TCP) is the most popular protocol
on the transport layer on the Internet protocol suite. There is an
agreed assumption that TCP remains to be the main transport protocol
on the Internet, thus it is the most targeted when interfering with the
privacy and anonymity of the users. A TCP header contains (among other
data) the source port and destination port. This information makes it
possible to identify (in most cases) what type of application
is being used over TCP; therefore allowing it to determine the kind of
inspection required to capture the data transmitted in the TCP
segment. 

Despite, the fact that some applications can use TCP ports out of range
of well-known ports, most of the Internet traffic is based on
protocols such as HTTP, SMTP, POP3, IMAP, etc; for which TCP ports and
protocol states are easily recognized while using DPI
techniques/solutions. Therefore, this facilitates the task of targeting
a particular network traffic and its processing for further
analysis. The biggest concerns, in terms of privacy and anonymity, is
that the current technology allows inspection of every single TCP segment
and its content. Despite encryption gaining in popularity, in both the
implementation of application protocols and as user practice, most of
the Internet traffic is still unencrypted. Hence, any traffic analysis
that is performed over the transport layer on unencrypted data
reveals the content of the transmission, thus, eliminating any kind of
privacy. 

In addition, due to TCP's architecture, the inspection
realized over the data flows is granular and accurate, mainly due to
the possibility to identify sources and destinations, together with
the content of the payload. There is still a possibility in which the
transport layer gets compromised but not the identity of the source
and destinatary. Nevertheless, an observer can gather enough
information to conclude over time, the final identity of the parties
involved, in the case that the payload contains sensitive data about their
identity.

\subsection{Application Layer}

The DPI solutions utilized over the application layer are varied, and
sometimes protocol specific\cite{taxo}. DPI techniques/technologies
focused on the application layer provide the possibility to capture
and recompose the data of the transmission. This generates high-risk
threats of the privacy and anonymity of the users exposing the
data of a particular application that the user is using. Examples include
the indiscriminate analysis of the web traffic that some
entities are performing\cite{taxo}; resulting in the compilation
of user profiles, type of content visiting, frequency, location,
etc. The possibility to cross-reference the analysis of different
application protocols utilized for the same user; building user
profiles that contain all kinds of personal and behavioral information
that gets stored and categorized without the user's consent.

Due to the fact that it is in this layer where the data gets originated, other
techniques get in place that can affect the privacy and
anonymity. Other popular protocols such as those dedicated to
e-mail, are common targets due to the sensibility of information that
they usually transmit. In this case, the threats against
privacy and anonymity does not necessarily happen in the transmission,
but in the software itself. Certain entities have implemented
control mechanisms over the software utilized to handle e-mail, such
as desktop clients and webmails.

Several  applications make heavy use of encryption
techniques in order to secure the communication on the layers
below. Nevertheless, there is a possibility that the methods utilized
by the application protocol to encrypt the communication are being
compromised even before the transmission begins. Examples of this include
possible backdoors recently claimed\cite{FS2007} on some of the
cryptographically secure pseudo-random number generators (CSPRNG)
utilized to generate encryption keys. If this is the case, certain
entities could decrypt traffic that is thought to be secure by the users,
leading again to an attack of their privacy and anonymity, even when
the users think they are using proper mechanisms to protect
themselves. These recent events open several questions about how
reliable a protocol can be if the cryptographic assumptions utilized
are somewhat misleading.

\section{Mechanism to enforce better privacy and anonymity}

In the previous section we enumerate the different technologies utilized
in order to establish certain controls over the privacy and anonymity
of the users. As it is common when any mechanism of oppression is enforced, 
several alternatives are created in order to avoid those mechanisms. Privacy
and anonymity have been a big concern since the beginning of the Internet.
In addition to those entities that are aiming to establish the above mentioned controls, there are also several companies, organizations and individuals, that are spending time and resources developing mechanisms of defense against the mentioned threats. We introduce the state of art concerning the enforcement of privacy and anonymity on the Internet.

\subsection{Technological principles behind the privacy and anonymity}

The majority of the techniques utilized to guarantee privacy are related to a combination of encryption and anonymity techniques. The vast majority of anonymity techniques rely on protecting the real identity through a combination of methods that are difficult to trace the origin and destination of the communication channel. Despite the complexity that encryption mechanisms can involve, most of the modern and popular application protocols provide the possibility to establish the connection through secure channels; either through the use of the Transport Secure Layer (TLS), or through the configuration of proxies or socket secure (SOCKS) mechanisms. 

There are certain methods to measure the grade of privacy and anonymity. The degree of privacy is mostly linked to the type of encryption utilized and computational capacity available. Different encryption algorithms are currently available, offering certain guarantees for the users. Several protocols in the application layer rely on these algorithms as the core of privacy enforcement. Some examples of this is the use of public-key cryptography\cite{rfc4880} and the use of algorithms such as RSA\cite{rfc3447} and DSA\cite{NIST}. In addition, and due to recent media revelations, some applications are moving to new cryptography schemes based on the use of elliptic curve cryptography such as Elliptic Curve Diffie-Hellman (ECDH), Integrated Encryption Scheme (IES) or  Elliptic Curve Digital Signature Algorithm (ECDSA). The main argument behind the use of new cryptography schemes, is the suspected evidences concerning the pseudo-random number generators utilized for them, and the possibility of broken cryptography\cite{FS2007}. Furthermore, the possibility to encapsulate the connections through a SOCKS interface allows the use of routing techniques through anonymous networks, that are difficult to trace.

\subsection{Proxy server}

One of the oldest technologies used to enforce privacy and anonymity has been the use of proxy servers. Perhaps due to the simplicity in their functioning, together with their popularity in the early days of the Internet, proxy servers still are one of the main technologies in use when enforcing privacy and anonymity.
The function of a proxy server consists mainly in masking the client requests, providing a new identity, i.e. a different IP address possibly located in a different geographical location. There is a vast number of proxy servers publicly and privately currently available such as www.anonymizer.ru, provides integrated proxy solutions within the web browser. In addition, several lists of free proxies are published daily across the Internet\footnote{anonymousproxylists.net, fresh-proxy-list.net, free-proxy-server-list.com, freeproxy.ru, nntime.com, etc}, offering all kind of proxies located in different countries, with different levels of anonymity. Notwithstanding, proxys servers cannot be considered a reliable method to guarantee privacy and anonymity by definition. The main reason being that the proxy server knows the origin and destination of the requests, therefore, if it is compromised, it can expose the identity of the users behind its use\cite{averview}. Also, although the proxy server can keep the identity secret, there is no guarantee that the content of the requests is not being monitored. Therefore, proxy servers cannot guarantee any property related to plausible deniability and true anonymity/privacy, thus they should be avoided as method to guarantee privacy and anonymity as defined in this paper.

\subsection{Onion routing}

\textit{Onion Routing is a general purpose infrastructure for private communication
over a public network\cite{ONION}}. The core architecture of onion routing is the implementation of mixed networks, i.e. nodes that are accepting messages from different sources and routing them randomly to other nodes within the network. The messages transmitted between them are encrypted, and different layers of encryption get removed in the process of routing while traveling across the nodes\cite{ONION}. This increases the difficulty of monitoring the traffic (thus revealing the identity). In addition, onion routing is not node-dependent, therefore, the compromising of a router does not compromise the network itself (although can facilitate the traffic analysis). Nevertheless, onion routing has some weaknesses while protecting the privacy and anonymity of its users.

Recent research\cite{SWARM} shows that even if the traffic is encrypted and hard-traceable, there are still possibilities to conclude the population of users and their geographical location. Furthermore, intersection attacks and timing analysis\cite{timing}, can reveal user's behavior within the onion routing network, leading to further identification of the sources or destinations. Despite these issues, onion routing is still considered one of the best alternatives when aiming to guarantee privacy and anonymity.

\subsection{TOR}

TOR, previously known as The Onion Router\cite{TOR}, is among the most popular solutions used these days to protect privacy and anonymity on the Internet. The main goal of TOR it is to provide a circuit-based low-latency anonymous communication service\cite{TOR}.TOR's core architecture is based on the same principles as onion routing. TOR contains several improvements over traditional onion routing, including: \textit{``perfect forward secrecy, congestion control, directory servers, integrity checking, configurable exit policies, and a practical design for location-hidden services via rendezvous points"}\cite{TOR}. Some nodes within the TOR network act as discovery servers, providing ``trusted" known routers, that are available for the end users. TOR routers have different roles and they can be classified as:

\begin{itemize}
\item {\textit{Middle-relay}: receive traffic and passes it to another relay}
\item {\textit{Bridges}: publicly listed and their main goal is to provide an entry point on networks under heavy surveillance and censorship}
\item {\textit{Exit-relay}: the final relay before reaching the destination and are publicly advertised}
\end{itemize}

\begin{figure}[h]
  \centering
  \captionsetup{justification=centering}
  \includegraphics[width=8cm]{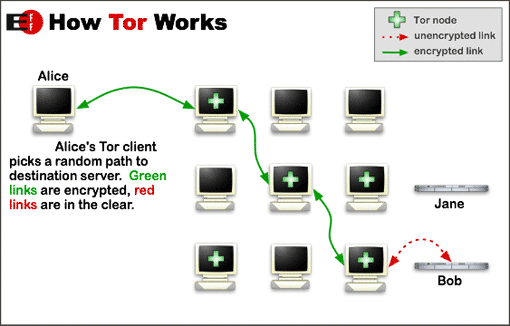}
  \caption{TOR functioning. \textit{(Source: EFF)}}
\end{figure}

TOR has won popularity across the most common Internet users. The main reason (in addition to its enforcement of privacy and anonymity) is due to the fact that the it is free-software and there are multiple cross-platform clients available. In addition, several extensions/add-ons are available for the most popular web-browsers, making TOR a very suitable solution. In addition, because TOR is configurable in most of the applications through a SOCK interface, TOR can be used for a broad number of protocols, facilitating anonymity in different type of services\cite{TOR}.

The success of TOR does not rely only on its core technology and principles, but on the network of volunteers that maintain the nodes. Due to TOR's design, anyone with enough bandwidth can provide a new router, allowing this to expand TOR worldwide. Nowadays, TOR network is composed of more than 5000 routers.

\begin{figure}[h]
  \centering
  \captionsetup{justification=centering}
  \includegraphics[width=9cm]{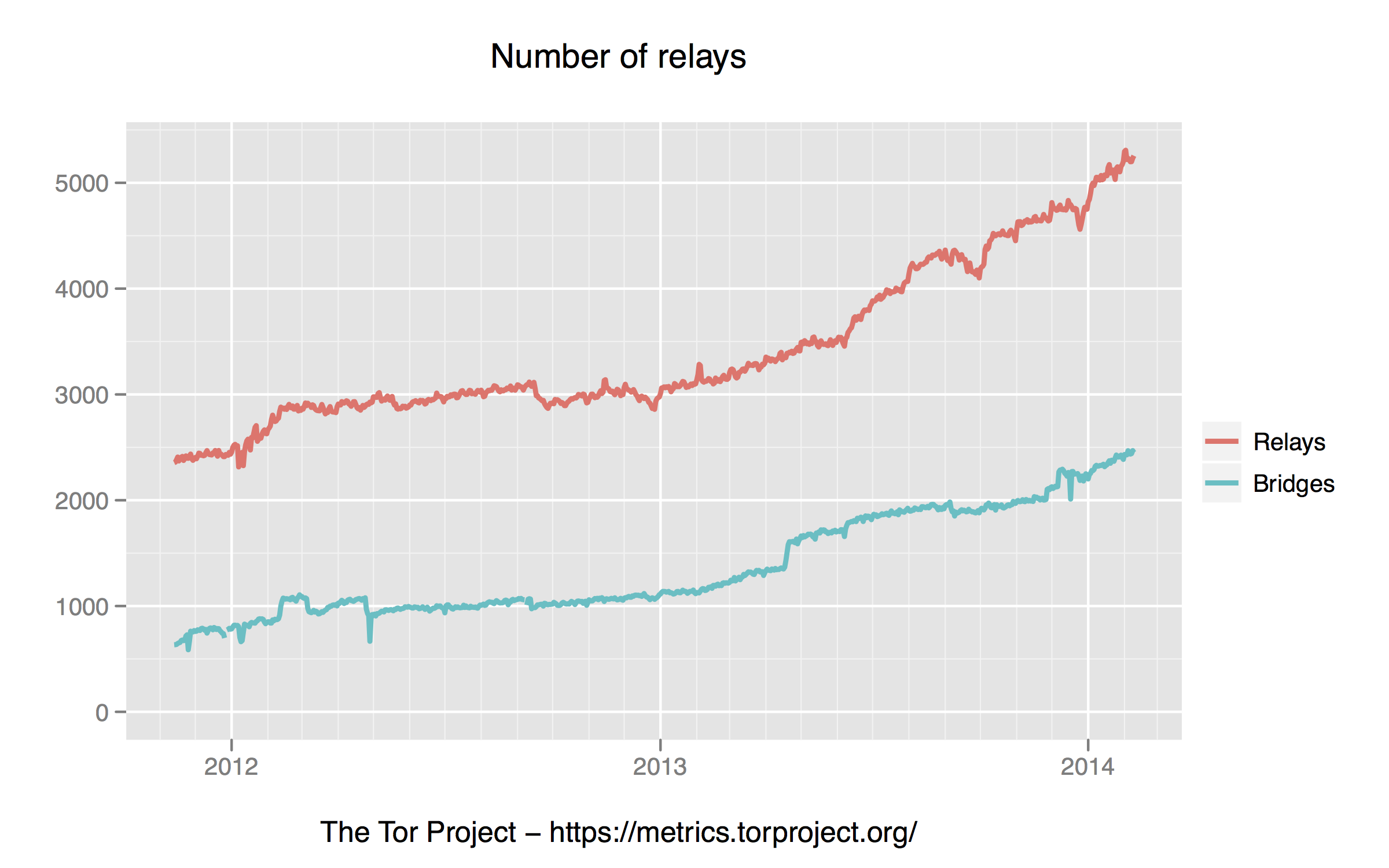}
  \caption{TOR relays \& bridges available}
\end{figure}

Furthermore, TOR user's population has been fluctuating over time. Recent media revelations concerning global surveillance programs, allowing TOR to win even more popularity across common users, making the network grow considerable, both on relays and users, in the past months, reaching peaks of more than \textbf{5 million} daily connections to the network.

\begin{figure}[h]
  \centering
  \captionsetup{justification=centering}
  \includegraphics[width=9cm]{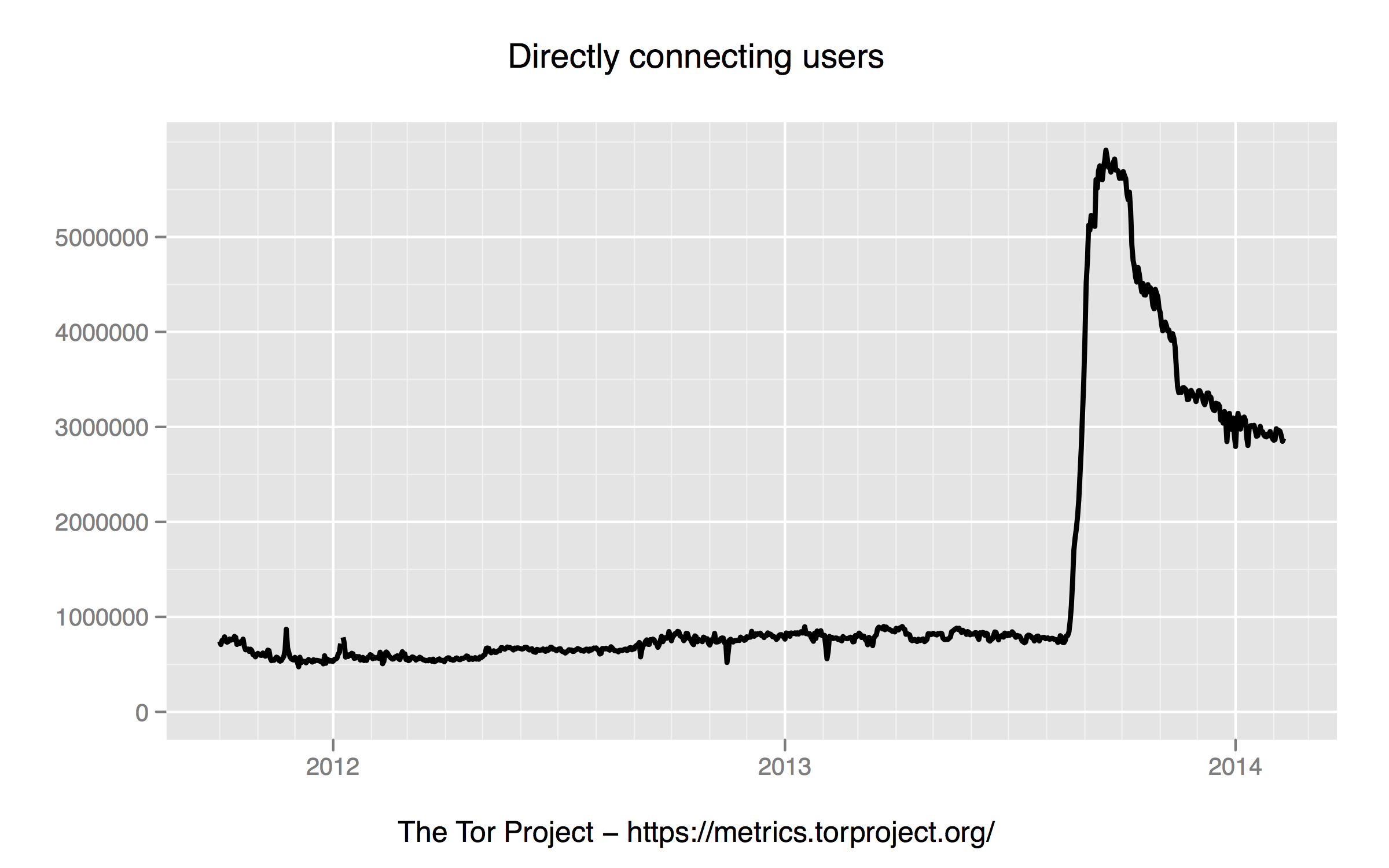}
  \caption{TOR user population}
\end{figure}

Despite its popularity, TOR is subjected to a certain degree of the techniques mentioned in Section 3. TOR architecture does not prevent statistical analysis over its traffic. Some authors highlight the possibility to interfere in nodes with high traffic, being able to perform traffic analysis with enough accuracy that can affect the anonymity of the network\cite{torattack}. In addition, and due to the publicity of the exit-relays, these particular nodes can be targeted either for traffic interception or Denial of Service  attacks (DoS), affecting the functionality of the network. Finally, TOR relay advertisement leads to the generation of blacklists by ISPs and governmental organizations that are aiming to control mechanisms enforcing privacy and anonymity, making it difficult for the end users to use TOR. In Section 5 we introduce the possible improvements to do in order to comply with the definitions stated in this paper.

\section{Related work and future improvements}

In the previous sections, we have analyzed the different motivations, technologies and state of the art concerning the threats and enforcement of privacy and anonymity on the Internet. All of the research available these days focuses mainly on two aspects related to privacy and anonymity: how to obtain secure and trustable encryption methods, and how to guarantee, real anonymous communications.

As of today, there is still no single solution available that is complaint with the definitions of privacy and anonymity introduced in Section 1. Nowadays, every communication done through the Internet that aims to be private and anonymous requires the use of cryptography and hard-trace routing techniques.  Proxies and TOR are among the most used anonymity technologies\cite{averview} and public-key cryptography is the most popular method aiming for privacy. Nevertheless, it is still possible to perform different attacks over these techniques, leading always to some result that invalidate the required properties for privacy and anonymity mentioned in Section 1. We introduce some of the possible improvements that current lines of research are proposing in order to enforce better privacy and anonymity.

\subsection{Possible solutions for traffic analysis}

Traffic analysis (both active and passive), is one of the primary techniques when targeting networks such as TOR. Recent research suggests that one way to achieve better privacy and anonymity over this kind of networks, will be the generation of random connections to achieve better plausible deniability\cite{SWARM}. In addition, there are some proposals to obfuscate TOR's traffic, adding delays or artificial traffic, together with package batching techniques\cite{AQUA}. Theoretically, these two improvements, will guarantee true anonymity by making the traffic analysis hard-computable. Nevertheless, there is no empirical evidence yet, mainly because they have not been implemented within the production network (i.e. with a real population of users).

\subsection{Blacklisting techniques}

Both proxy servers and TOR routers are suffering blacklisting techniques. Because the implicit principles of both technologies (available public nodes), it will require a modification of how they are used and discovered, in order to avoid blacklisting. The main proposals aim for using a rendezvous protocol as the only way to distribute the address of the nodes\cite{boost}

\subsection{Better encryption methods}

There is a common effort within the cryptography community, to provide more robust and reliable encryption schemes. In the last year, there has been new contributions related to Elliptic curve cryptography, specially those not linked to any corporation or standardization organization such as the Curve25519\cite{25519}. It is expected that several well known protocols will start using  these new cryptography schemes, thus avoiding the possible backdoors established in the current/popular schemes utilized.

\subsection{ISP cooperation}

It has been highlighted by some authors\cite{cirri}, that ISPs are playing a crucial role in privacy and anonymity matters. They are the main entities responsible for implanting mechanisms that can affect the privacy and anonymity of the users. Nevertheless, there are not too many ``friendly" ISPs in these matters. Some proposals involve the commitment of the ISP to provide better quality relay-bridges for TOR\cite{cirri}. Also, there have been official requests to get more neutrality and protection from the ISPs. Nevertheless, all of them are still subjected to the laws that sometimes conflict with the mechanisms for guaranteeing privacy and anonymity.

\section{Conclusion}

In this paper we introduced the definitions of privacy and anonymity according to the latest related standards of the field. We analyzed the main originators behind the threats against the privacy and anonymity on the Internet; we enumerate their motivations, together with the techniques and technologies utilized to control the privacy and anonymity on the Internet. In addition, we gave an overview of the state of the art concerning the enforcement of privacy and anonymity on the Internet, enumerating the most common technologies by the users, together with their shortcomings. Finally, we introduced the new lines of research together with their proposals. We conclude that there is not yet available any method that guarantees real privacy and true anonymity.

The author considers that despite the reasons given by governments and corporations, privacy and anonymity are a human right that must be preserved no matter the channel of communication used. It is a moral responsibility for the scientific community to research, develop and implement better technologies to guarantee our fundamental rights.

\bibliography{mybib}{}
\bibliographystyle{plain}
\end{document}